# IDENTIFYING CURRICULUM DISRUPTIONS IN ENGINEERING EDUCATION THROUGH SERIOUS GAMING

**Roger Waldeck**[a,], **Ann-Kristin Winkens**[b,1], **Clara Lemke**[c],
**Carmen Leicht-Scholten**[d], **Haraldur Audunsson**[e]

[a] IMT Atlantique, Brest, France, ORCID 0000-0003-2031-9054
[b] RWTH Aachen University, Aachen, Germany, ORCID 0000-0003-4637-3905
[c] RWTH Aachen University, Aachen, Germany, ORCID 0009-0003-4623-6019
[d] RWTH Aachen University, Aachen, Germany, ORCID 0000-0003-2451-6629
[e] Reykjavik University, Reykjavik, Iceland, ORCID 0000-0002-7730-346X

## ABSTRACT

This workshop introduces participants to SUCRE, a serious game designed to enhance curriculum resilience in higher education by simulating crisis scenarios. While applicable to various disciplines, this session focuses on engineering curricula, identifying discipline-specific challenges and potential adaptations. Participants will engage in Step 1 of the game, analyzing trigger events and their impacts on curriculum structures. At the end of the workshop, attendees will be able to identify key triggers that may affect curricula, assess their cascading effects, and reflect on the applicability of SUCRE within their own institutions.

## 1 BACKGROUND

### 1.1 Navigating VUCA challenges in Higher Education

Crises and extreme events are increasingly prevalent (Hällgren et al., 2018). The complexity of globally connected socio-technical systems in which higher education institutions (HEIs) operate presents new challenges in preparing for, responding and adapting to a rapidly changing VUCA (volatility, uncertainty, complexity, ambiguity) world (Kamp, 2020). VUCA situations are characterized by intricate cause-and-effect relationships, environmental volatility, high uncertainty due to incomplete information, and multiple, often conflicting interpretations among stakeholders (Waldeck et al., 2019). Unlike traditional risk management, resilience acknowledges that some disruptions are unforeseeable, yet organizations can still prepare for them. The Covid-19 crisis highlighted significant differences in HEI's resilience profiles (Berthoud et al., 2021; McKeown et al., 2022). Those who exhibited weaker performance, must now endeavor to learn from the experience and enhance their preparedness for future challenges (Habersaat et al., 2020). Technical universities and engineering education programs, in particular, seem to be highly vulnerable to sudden disruptions due to their frequent reliance on traditional teaching methods and rigid curriculum structures, which often lack the flexibility needed to adapt to rapidly changing conditions (Brink et al., 2024; Hadgraft & Kolmos, 2020).

---

[1] Corresponding Author
A Winkens
ann-kristin.winkens@rwth-aachen.de

Building resilience requires a comprehensive strategy that defines the purpose for resilience (i.e., resilience for what), identifies potential disruptions to deal with (resilience to what), determines internal and external sources of resilience (the how) and specifies when and which resilience capacities should be activated (the when) (Duchek, 2020). Some scholars define resilience as the ability to maintain essential operations (Hollnagel, 2010) or to rapidly restore pre-crisis conditions (Holling, 1973). More recently, a resilience concept has emerged, where organizations leverage adversity to develop an improved operational model (Chen et al., 2021; Macrae & Wiig, 2019).

Serious games (SG) can serve as valuable tools for HEIs, supporting effective decision-making in complex scenarios. SG are increasingly used to support long-term learning and facilitate decision-making for complex challenges, such as disaster risk management and sustainable socio-economic systems (Taillandier & Adam, 2018; van Schaik, 2023). By increasing awareness of unforeseen events and their potential consequences, SGs help institutions to strengthen curriculum resilience and better prepare for future disruptions, thereby achieving an enhanced resilience level for their HEI curriculum.

## 1.2 A Serious Game for Curriculum Design

The DECART (Agility, Resilience, and Transformation in Curriculum Design) project fosters the exchange of innovative curricula and strategies for implementation in VUCA contexts. It supports curriculum cooperation, leadership development, and dissemination of advanced curriculum design methods. DECART is an international collaboration involving institutions from France, Germany, Iceland, Indonesia, Lithuania, and South Africa, which have a particular focus on engineering and business education.
As part of DECART, a serious game ("SUCRE" for Serious University Curriculum REsilience game) was designed to enhance the resilience of curricula in higher education institutions. Within DECART, the term "curriculum" is generally understood as guidance on what to teach and the educational process leading to specific learning outcomes (Matthiasdottir et al., 2024).
In the SUCRE serious game, players analyze and respond to trigger events that may disrupt higher education curricula. These VUCA events (Waldeck et al., 2019) generate multiple interpretations and evaluations, reflecting the complexity of real-world decision-making. The VUCA situation is highlighted in the game through different aspects: complexity arises from the interdependencies between a trigger and its consequences (Figure 1), uncertainty stems from the challenge of determining whether the trigger's impact will be positive or negative, volatility is reflected in the rapid and unpredictable changes certain trigger events can provoke, which can create sudden disruptions in curriculum structures, and ambiguity becomes apparent in the analysis of which consequences to consider, as well as the diverse interpretations participants may have.

Players work in teams to strengthen a virtual university's curriculum resilience. In SUCRE, following the resilience construct of McManus et al. (2008), three capacities are to be acquired and applied through playing the game: 1) building situational awareness by understanding the ripple and feedback effects of a given trigger event through consequence scenario analysis, 2) identifying vulnerabilities of a higher education institution curriculum to a given trigger event and prioritizing the events to focus on, and 3) building adaptive capacity by designing transformative actions in response to these challenging triggers, based on the insights gained from steps 1) and 2).

Trigger and impact cards enhance situational awareness for the players. Trigger cards depict crises or changes affecting higher education, categorized into (i) competition and economy, (ii) political and environmental context, and (iii) technology and innovation. Impact cards outline potential consequences, which can be interconnected and are affected either positively, negatively or in an undetermined or non-quantifiable way. Recognizing triggers, assessing consequences using impact cards, and developing a shared understanding are central to resilience. Figure 1 illustrates how players could use consequence scenario mapping to enhance situational awareness. Here an exemplary trigger card "education costs will increase significantly for students" generates cascading impacts.

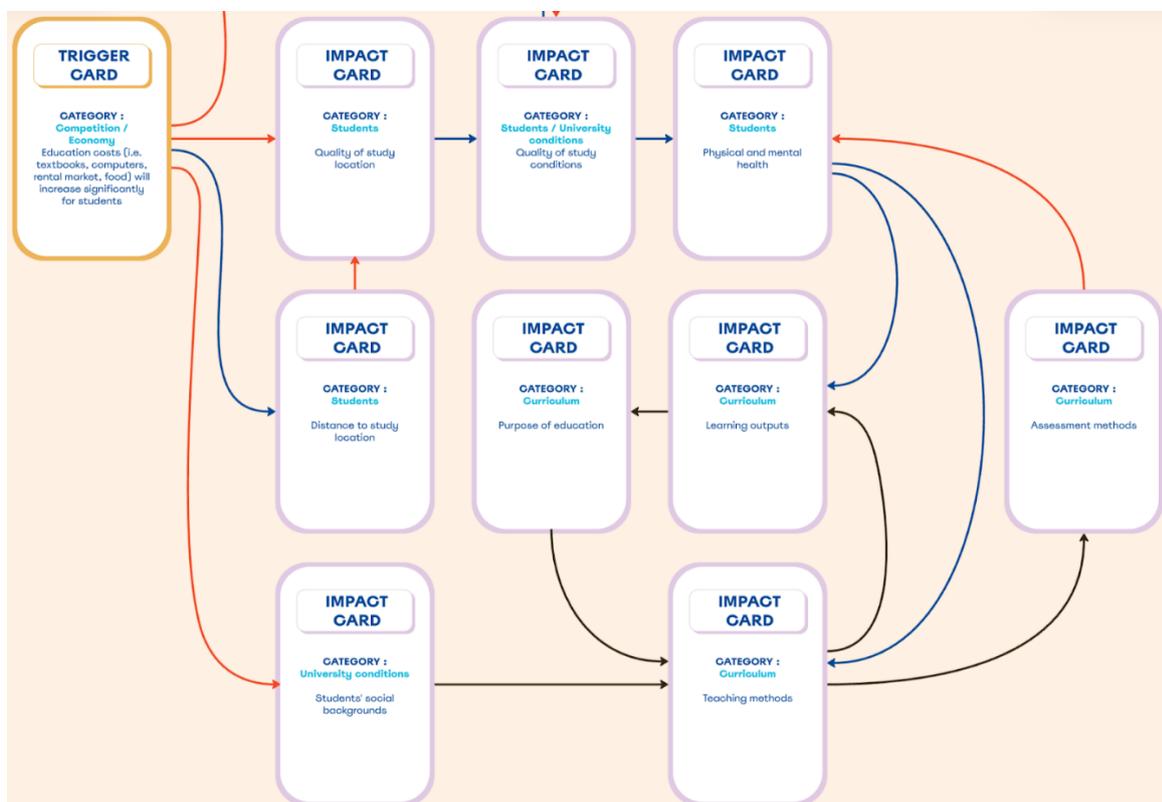

*Figure 1. An example from the game showing the direct (impact cards) and cascading effects from one VUCA event (trigger card). The color of each arrow indicates if the effect is positive, negative or indeterminate.*

Curriculum Identity cards (ID) in SUCRE represent different types of curricula from different universities, each with unique strengths and vulnerabilities. Players receive a curriculum ID card describing a university and its curriculum along various dimensions.

## 2 WORKSHOP OBJECTIVES

This workshop introduces SUCRE and its application to engineering curricula. While SUCRE is applicable to various disciplines and institutional contexts, this workshop aims to explicitly focus on engineering curricula as a use case, exploring discipline-specific challenges and potential adaptations.

The core objective of the workshop is to engage participants in Step 1 of the game called "**building situational awareness**" among participants: **Identifying trigger events and analyzing their potential impacts on engineering curricula (**as it is exemplarily shown in Figure 1).

### 2.1 Expected learning outcomes

By the end of the workshop, participants will:
- Identify key trigger events affecting engineering curricula, such as technological disruptions, regulatory changes, and funding shifts.
- Analyze ripple effects and feedback loops using structured scenario mapping.
- Reflect on the applicability of SUCRE within participants' institutions and discuss discipline-specific considerations for engineering education.

### 2.2 Target Audience

This workshop is designed for:
- Engineering educators and program leaders interested in preparing curricula for unexpected disruptions.
- Higher education administrators working on curriculum development.
- Researchers in engineering education and curriculum innovation exploring new approaches to adaptive learning structures.

No prior experience with SUCRE or serious gaming is required.

## 3 WORKSHOP DESIGN

This interactive session combines game-based scenario analysis, structured discussions, and reflection. Participants will work in small groups to assess crisis scenarios affecting an engineering curriculum. In this board game, they are provided with an Identity Card, Trigger and Impacts Cards, as well as a template that is used for the analysis.

**Time Plan (60 minutes)**
1. Introduction and Context (10 min)
    - Brief overview of curriculum resilience
    - Introduction to SUCRE and workshop objectives
2. Game Activity – Identifying Triggers and Impacts (35 min)

- Each group is assigned a fictional engineering program/university (ID card)
- Each group analyzes given potential triggers, mapping direct and cascading impacts on the curriculum.
- Groups discuss and document key vulnerabilities and findings.
3. Discussion and Takeaways (15 min)
    - Groups share insights and compare interpretations of the trigger events.
    - Open discussion on engineering-specific curriculum challenges and potential game adaptations.
    - Reflection on how participants can apply these insights in their own institutions.

By focusing on engineering education, the workshop will allow for discipline-specific insights and discussions on potential customizations of SUCRE for this field. Participants will leave with a deeper understanding of curriculum vulnerabilities and a framework for analyzing and responding to disruptions in engineering education.

## 4   WORKSHOP RESULTS

Two groups participated in the workshop with participants from the Netherlands, United Kingdom, France, Poland and Norway.
Both groups worked with the same Identity Card (Environmental Engineering) and receiving the same set of Trigger and Impact Cards. Interestingly, the groups approached the exercise differently and arrived at distinct outcomes, demonstrating the diversity of perspectives that may arise when reflecting on different triggers and impact on a curriculum.

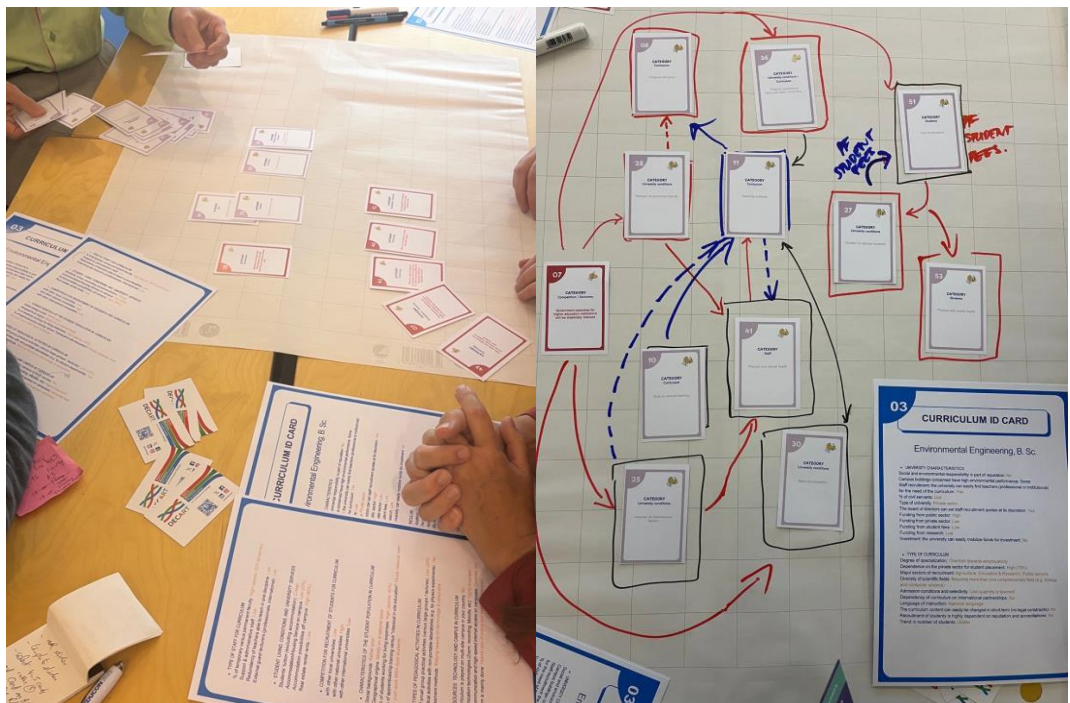

*Figure 2. Impressions from the SUCRE Game Workshop at SEFI.*

The first group focused on one main trigger card and decided to analyze its potential effects in depth. Their mapping emphasized mostly negative impacts. Beforehand, they screened possible triggers and then agreed on one specific event to explore. In their scenario mapping of consequences, several arrows indicated double or reinforcing effects, highlighting how impacts can amplify each other. The second group chose to integrate two trigger cards and also analyze mostly negative impacts.

In general, the activity showed that the game can be applied at different levels of detail by starting from a single trigger and its immediate impacts and then extending the map step by step. The participants appreciated this approach and engaged in a lively discussion, highlighting the value of the exercise. Discussions also revealed the importance of distinguishing between short-term and long-term impacts, as the same trigger may generate immediate disruption but also create opportunities for adaptation over time. Moreover, we focused only on the first step of the game due to the one-hour time limitation, which proved effective for the session format. At the same time, the exercise highlighted the importance of considering the diversity of different national higher education systems when reflecting on possible triggers and impacts on curricula. Finally, it became clear that the game can be played both within a shared educational setting (e.g., participants from the same university working on a specific curriculum) and among participants from different institutions, programs, or national contexts, with both constellations offering valuable perspectives.

For future applications, it may be useful to provide both a basic version of the game with fewer cards for a faster entry and an extended version with a larger card set for deeper exploration.

## REFERENCES

Berthoud, L., Lancastle, S., & Gilbertson, M. (2021). Designing a Resilient Curriculum for a Joint Engineering First Year. In *Proceedings of the 49th Annual Conference of the European Society for Engineering Education* (pp. 664–676). https://doi.org/10.5281/zenodo.14647128

Brink, S. C., Miranda, d. H., Ellen, S., Johan, C. C., Fredrik, G., Elizabeth, K., . . . and Admiraal, W. (2024). Curriculum Agility principles for transformative innovation in engineering education. *European Journal of Engineering Education*, 1-17. https://doi.org/10.1080/03043797.2024.2398165

Chen, R., Xie, Y., & Liu, Y. (2021). Defining, Conceptualizing, and Measuring Organizational Resilience: A Multiple Case Study. *Sustainability*, *13*(5), 2517. https://doi.org/10.3390/su13052517

Duchek, S. (2020). Organizational resilience: a capability-based conceptualization. *Business Research*, *13*(1), 215-246. https://doi.org/10.1007/s40685-019-0085-7

Habersaat, K. B., Betsch, C., Danchin, M., Sunstein, C. R., Böhm, R., Falk, A., . . . Butler, R. (2020). Ten considerations for effectively managing the COVID-19 transition. *Nature Human Behaviour*, *4*(7), 677-687. https://doi.org/10.1038/s41562-020-0906-x

Hadgraft, R. G., & Kolmos, A. (2020). Emerging learning environments in engineering education. *Australasian Journal of Engineering Education*, *25*(1), 3–16. https://doi.org/10.1080/22054952.2020.1713522

Hällgren, M., Rouleau, L., & de Rond, M. (2018). A Matter of Life or Death: How Extreme Context Research Matters for Management and Organization Studies. *12*(1), 111-153. https://doi.org/10.5465/annals.2016.0017

Holling, C. S. (1973). Resilience and Stability of Ecological Systems. *Annual Review of Ecology and Systematics*, *4*(1), 1-23. https://doi.org/10.1146/annurev.es.04.110173.000245

Hollnagel, E. (2010). How Resilient Is Your Organisation? An Introduction to the Resilience Analysis Grid (RAG). *Sustainable Transformation: Building a Resilient Organization*. https://minesparis-psl.hal.science/hal-00613986

Kamp, A. (2020). *Navigating the Landscape of Higher Engineering Education. Coping with decades of accelerating change ahead*. TU Delft.

Macrae, C., & Wiig, S. (2019). Resilience: From Practice to Theory and Back Again. In S. Wiig & B. Fahlbruch (Eds.), *Exploring Resilience: A Scientific Journey from Practice to Theory* (pp. 121-128). Springer International Publishing. https://doi.org/10.1007/978-3-030-03189-3_15

Matthiasdottir, A., Audunsson, H., Dagienè, V., Rouvrais, S., Barus, A., & Gerwel Proches, C. (2024). Examining Best Practices in Curriculum Design: Insights for Engineering Education. In *Proceedings of the 52nd Annual Conference of the European Society for Engineering Education* (pp. 727-736). https://doi.org/10.5281/zenodo.14254854

McKeown, J. S., Bista, K., & Chan, R. Y. (2022). COVID-19 and higher education: Challenges and success during the global pandemic. In J. S. McKeown, K. Bista, & R. Y. Chan (Eds.), *Global higher education during COVID-19: Policy, society, and technology* (pp. 1-7). STARS Scholars. https://ojed.org/cies/article/view/4199

McManus, S., Seville, E., Vargo, J., & Brunsdon, D. (2008). Facilitated Process for Improving Organizational Resilience. *Natural Hazards Review*, *9*(2), 81-90. https://doi.org/10.1061/(ASCE)1527-6988(2008)9:2(81)

Taillandier, F., & Adam, C. (2018). Games Ready to Use: A Serious Game for Teaching Natural Risk Management [Article]. *Simulation & Gaming*, *49*(4), 441-470. https://doi.org/10.1177/1046878118770217

van Schaik, F. (2023). What happens if …? Uncertainty in games and climate change education. *Environmental Education Research*, *29*(12), 1891-1910. https://doi.org/10.1080/13504622.2023.2225811

Waldeck, R., Gaultier Le Bris, S., & Rouvrais, S. (2019). Interdisciplinarity and VUCA. In *Methods and Interdisciplinarity* (pp. 99-116). https://doi.org/10.1002/9781119681519.ch5